\documentclass[letterpaper]{article} % DO NOT CHANGE THIS
\usepackage{aaai2026}  % DO NOT CHANGE THIS
\usepackage{times}  % DO NOT CHANGE THIS
\usepackage{helvet}  % DO NOT CHANGE THIS
\usepackage{courier}  % DO NOT CHANGE THIS
\usepackage[hyphens]{url}  % DO NOT CHANGE THIS
\usepackage{graphicx} % DO NOT CHANGE THIS
\usepackage{natbib}  % DO NOT CHANGE THIS AND DO NOT ADD ANY OPTIONS TO IT
\usepackage{caption} % DO NOT CHANGE THIS AND DO NOT ADD ANY OPTIONS TO IT
\usepackage{booktabs}
\usepackage{amsmath}
\usepackage{amssymb}
\usepackage{makecell}
\usepackage{algorithm}
\usepackage{algorithmic}

\usepackage{newfloat}
\usepackage{listings}
\DeclareCaptionStyle{ruled}{labelfont=normalfont,labelsep=colon,strut=off} % DO NOT CHANGE THIS
\floatstyle{ruled}
\newfloat{listing}{tb}{lst}{}
\floatname{listing}{Listing}
\title{Selection and Exploitation of High-Quality Knowledge from Large Language Models for Recommendation}
\author{
    Guanchen Wang\equalcontrib\textsuperscript{\rm 1}, Mingming Ha\equalcontrib\textsuperscript{\rm 1}, 
    Tianbao Ma\equalcontrib\textsuperscript{\rm 1}, \\
    Linxun Chen\textsuperscript{\rm 1}, Zhaojie Liu\textsuperscript{\rm 1}, Guorui Zhou\textsuperscript{\rm 1}, Kun Gai\textsuperscript{\rm 2}
}
\affiliations{
    \textsuperscript{\rm 1}Kuaishou Technology, Beijing, China\\
    \textsuperscript{\rm 2}Unaffiliated\\
    \{wangguanchen, hamingming, matianbao, chenxi36, zhaotianxing, zhouguorui\}@kuaishou.com,
    gaikun@qq.com
}

\usepackage{bibentry}
\begin{document}

\maketitle

\begin{abstract}
In recent years, there has been growing interest in leveraging the impressive generalization capabilities and reasoning ability of large language models (LLMs) to improve the performance of recommenders. With this operation, recommenders can access and learn the additional world knowledge and reasoning information via LLMs. However, in general, for different users and items, the world knowledge derived from LLMs suffers from issues of hallucination, content redundant, and information homogenization. Directly feeding the generated response embeddings into the recommendation model can lead to unavoidable performance deterioration. To address these challenges, we propose a Knowledge Selection \& Exploitation Recommendation (KSER) framework, which effectively select and extracts the high-quality knowledge from LLMs. The framework consists of two key components: a knowledge filtering module and a embedding spaces alignment module. In the knowledge filtering module, a Embedding Selection Filter Network (ESFNet) is designed to assign adaptive weights to different knowledge chunks in different knowledge fields. In the space alignment module, an attention-based architecture is proposed to align the semantic embeddings from LLMs with the feature space used to train the recommendation models. In addition, two training strategies--\textbf{all-parameters training} and \textbf{extractor-only training}--are proposed to flexibly adapt to different downstream tasks and application scenarios, where the extractor-only training strategy offers a novel perspective on knowledge-augmented recommendation. Experimental results validate the necessity and effectiveness of both the knowledge filtering and alignment modules, and further demonstrate the efficiency and effectiveness of the extractor-only training strategy.
\end{abstract}


\section{Introduction}
Recently, the rapid development large language models (LLMs) has brought profound impacts across various fields. Exemplified by architectures like GPT \cite{radford2018improving}, BERT \cite{BERT}, encoder-decoder Transformer \cite{NIPS2017_3f5ee243}, and others, the impressive language understanding, generation and reasoning performance of LLMs allows them to effectively adapt to a wide range of unseen tasks and domains. Especially for recommendation tasks, the abilities in sequence modeling and general-purpose reasoning demonstrate great potential to enhance and even revolutionize existing recommender systems \cite{RethinkingCDS,xu2023neural,xu2024slmrec,RecInLLMEra}. A natural idea is to directly or indirectly leverage these abilities of LLMs to promote the development of recommender systems. In the integration of LLMs and recommender systems, there currently exist three representative paradigms: LLMs serve as recommenders, LLM-based autonomous agent, bridge LLMs and recommendation models \cite{RecInLLMEra}. LLMs serving as recommenders, is also known as the generative recommendation, which directly performs recommendation tasks and predict the next user-interacted item. Some promising works have made some progress in this direction. A self-attention based sequential model (SASRec) \cite{SASRec} is designed to balance model complexity and expressiveness by behavior patterns while making predictions based on relevant past interactions. Inspired by the success of Transformer architecture, HSTU \cite{ActionsSpeakLouder} generative recommender is developed to address large-scale, high-cardinality, and non-stationary streaming recommendation tasks. Some generative recommenders such as Tiger \cite{Tiger}, OneRec \cite{OneRec}, etc., largely reduce the vocabulary size by transforming items into multiple semantic IDs with various approaches. For LLM-based autonomous agent, the core idea is to enable LLMs to simulate human-level user behaviors within recommender systems \cite{RecInLLMEra,RecAIAgent,UserBehaviorSim}. For the paradigm of bridging LLMs and recommendation models, \cite{TowardsOpen-WorldRec} achieves knowledge augmentation and compatibility for the traditional recommendation models by a mixture-of-expert-based adaptor. \cite{R4ec} iteratively refines the output responses of LLMs by the proposed reasoning, reflection and refinement framework, which interacts with the LLM and provide timely feedback to improve the response quality.

In current industrial applications, the majority of deployed recommender systems are still discriminative models. Generative recommenders require substantial reengineering of existing online recommendation models and even the entire recommendation pipeline, posing significant challenges to the stability and computational efficiency of online recommendation services. Therefore, this paper focuses on effectively leveraging the world knowledge and reasoning abilities of LLMs, which aims to bridge the gap between responses from LLM and recommendation models to provide data augmentation for training recommendation models. However, the world and reasoning knowledge from LLMs often contains redundant and highly homogeneous text contents, where only a limited portion constitutes high-quality and valuable knowledge for training recommendation models. Naively integrating the textual embeddings LLM alongside traditional recommendation features may lead to the model performance degradation. In the training process, the recommendation model may fail to effectively identify and leverage the most informative knowledge. Therefore, it is necessary to design a framework to achieve the selection and exploitation of high-quality knowledge from LLMs for improving the performance of the existing recommendation models with minor changes to the training pipeline. In this work, we develop a knowledge selection and embedding alignment framework with two training strategies to achieve this objective. The main contributions of this paper are summarized as follows
\begin{itemize}
    \item A Knowledge Selection \& Exploitation Recommendation (KSER) framework is proposed, which explicitly selects high-quality knowledge from LLMs, filters out redundant and negative knowledge, and then achieves the space alignment between semantic embeddings and traditional feature embeddings.
    \item Two specific model architectures are designed, serving as the knowledge selection and filtering (KSF) and embedding spaces alignment (ESA), respectively.
    \item Two training strategies are proposed to adapt different application scenarios and can be seamlessly integrated into existing recommender systems, where the extractor-only training strategy offers a novel perspective on knowledge-augmented recommendation. 
    \item Extensive experiments are conducted to demonstrate the performance of both the proposed KSER framework and the training strategies. The experimental results also verify the necessity and effectiveness of both the knowledge filtering and the embedding spaces alignment modules.
\end{itemize}

\section{Related Works}
To date, a large body of research has emerged on advancing the field of generative recommendation and knowledge-augmented recommendation. Generative recommendation has made significant progress in recent years. Several representative methods such as HSTU \cite{ActionsSpeakLouder}, Tiger \cite{Tiger}, OneRec \cite{OneRec}, MTGR \cite{MTGR} etc., are developed to reshapes the recommendation systems through end-to-end generative schemes. In \cite{Tiger}, a transformer-based generative retrieval approach is proposed, in which a semantically meaningful tuple of codewords is creatively constructed to serve as a semantic ID for each item. \cite{MTGR} proposes MTGR, a generative recommendation approach based on HSTU mentioned in Introduction, which preserves critical cross features from traditional recommendation models and enables efficient scaling through user-level compression. On the other hand, some approaches still employ discriminative recommendation models as the backbone of the recommender systems, using LLM outputs to assist and enhance the performance of the existing discriminative model. In \cite{TowardsOpen-WorldRec}, a knowledge augmented recommendation (KAR) framework is proposed, which introduces a mixture of expert architecture to integrate the reasoning knowledge and the factual knowledge into the existing recommendation models. \cite{BridgUser} proposes a hybrid recommendation framework that constructs a collaborative interest knowledge graph to improve recommendation performance by integrating inferred user interests from LLMs with item-side and collaborative information.

\section{Preliminaries}
\begin{figure*}[t]
\centering
\includegraphics[width=0.9\textwidth]{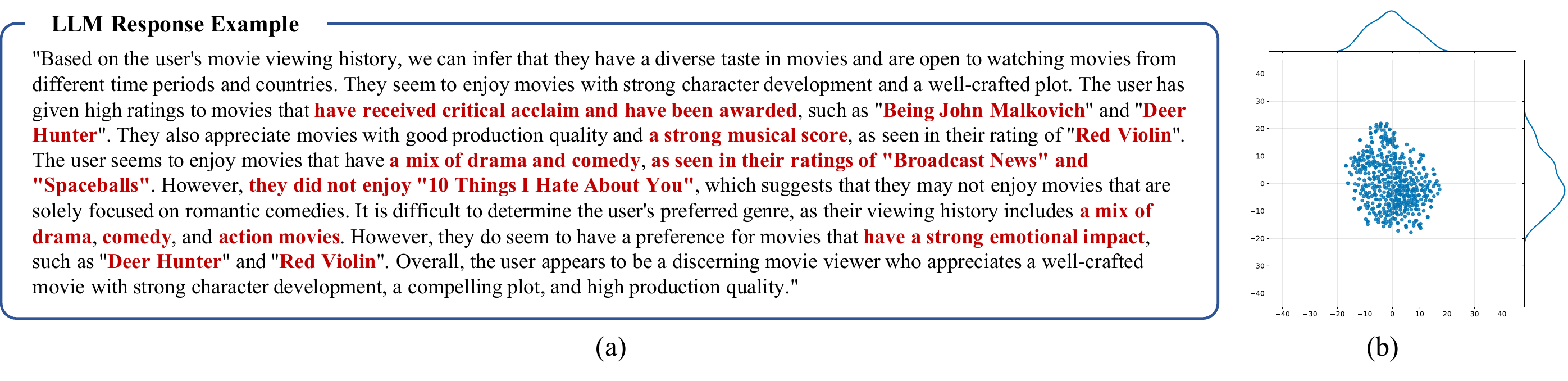} 
\caption{Content redundancy \& homogenization. (a) Redundant information in LLM responses; (b) The phenomenon of response homogenization.}
\label{motivation}
\end{figure*}

Consider a class of discriminative recommendation tasks, such as rating, click-through rate (CTR) and post-click conversion rate (CVR) predictions, and so forth, which are typically formulated as a supervised learning problem. The traditional recommendation dataset is defined as 
$\textbf{D}=\{(\textbf{x}_1, y_1),\dotsc,(\textbf{x}_i, y_i),\dotsc,(\textbf{x}_N, y_N)\}$, where $\textbf{x}_i=\big[\boldsymbol{x}_i^{(1)},\boldsymbol{x}_i^{(2)},\dotsc,\boldsymbol{x}_i^{(F)}\big]$ with $F$ feature fields, $y_i\in\{0,1\}$ corresponding to a binary classification problem or $y_i\in\mathbb{R}$ for a regression problem is the label of the $i$-th sample, $N$ is the number of data points. For CTR and CVR prediction tasks, the core objective is to establish a model to predict the click or conversion probability $\text{Pr}(y_i=1\vert\textbf{x}_i)$. On the other hand, the reasoning and world knowledge from LLMs is introduced to provide data augmentation for training recommendation models by a text encoder, e.g., BERT \cite{BERT}, Qwen3 Embedding series \cite{Qwen3Embed}, which forms a knowledge repository $\textbf{K}=\{\textbf{k}_1,\dotsc,\textbf{k}_i,\dotsc,\textbf{k}_N\}$. The encoding vector $\textbf{k}_i=\big[\boldsymbol{k}_i^{(1)},\boldsymbol{k}_i^{(2)},\dotsc,\boldsymbol{k}_i^{(L)}\big]$ denotes the world knowledge embedding with $L$ fields for the $i$-th sample, where $\boldsymbol{k}_i^{(j)}\in\mathbb{R}^{d_k}$ is a $d_k$-dimensional embedding. For recommendation models, the knowledge repository $\textbf{K}$ generally contains an amount of homogeneous and redundant information, and even includes some erroneous information, leading to deterioration of the recommendation performance. As shown in Fig. \ref{motivation}(a), the red words represent the information that is beneficial to the recommendation models. While the other words contribute to text fluency, they do not enhance the model’s performance and instead increase the modeling burden on the recommender system when extracting useful information from LLM outputs. Besides, the 1000 randomly selected textual embedding of responses from LLM after T-SNE  are given in Fig. \ref{motivation}(b), which demonstrate that the content homogenization results in limited discriminability among textual embeddings.
On the other hand, semantic embeddings of the knowledge from the text encoder and the feature fields used in recommenders typically require adaptation and alignment. Therefore, this work focuses on selection and exploitation of the high-quality knowledge generated by LLMs to enhance recommender systems.
\section{Methodology}
In this section, the Knowledge Selection \& Exploitation Recommendation (KSER) framework with two training strategies is formulated. 
\subsection{KSER Framework}
\begin{figure*}[t]
\centering
\includegraphics[width=\textwidth]{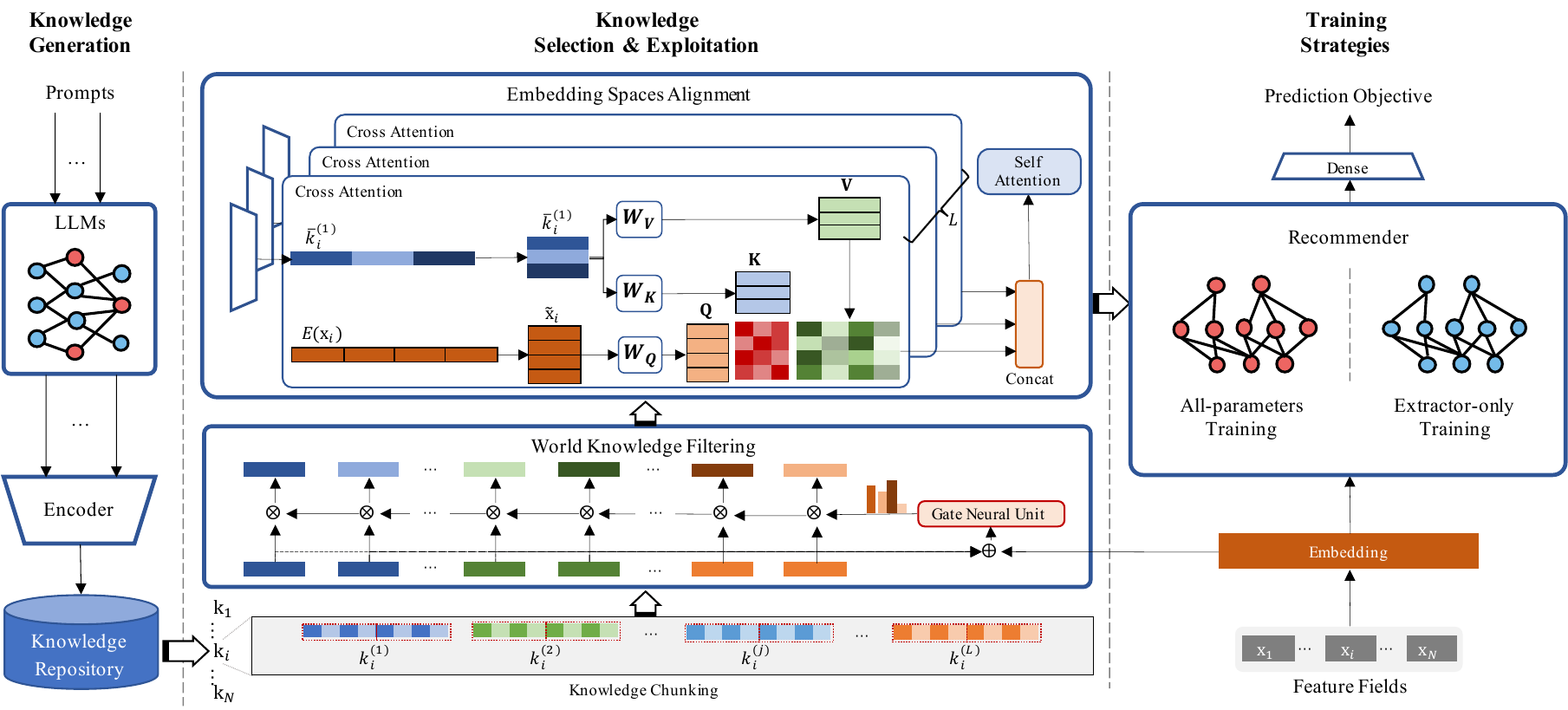} % Reduce the figure size so that it is slightly narrower than the column.
\caption{The overall framework of the proposed KSER with two training strategies. Embedding vectors in each knowledge field from the knowledge repository are first partitioned into several vector chunks. These vector chunks are selected and filter by the designed ESFNet. Then, the selected knowledge embeddings are aligned with the feature embeddings used in the recommendation models. Then the obtained output is fed into the recommenders.}
\label{Framework}
\end{figure*}

The world knowledge can be obtained from LLMs by designing prompts. Prompt design is not the focus of this work. Instead, we concentrate on selecting and exploiting high-quality textual knowledge and information from LLMs. As shown in Fig. \ref{Framework}, different types of prompts can be designed to generate different textual knowledge from multiple perspectives, such as scenario-side reasoning prompt, user preference reasoning prompt and item factual prompt mentioned in \cite{TowardsOpen-WorldRec}, etc.  
The textual knowledge from LLMs is subsequently transformed into knowledge representations $\textbf{K}=\{\textbf{k}_1,\dotsc,\textbf{k}_i,\dotsc,\textbf{k}_N\}$ through a text encoder, which constitutes an external knowledge repository for recommendation. The KSER framework consists of two core components designed to select and extract high-quality knowledge from the knowledge repository, namely the knowledge selection and filtering (KSF) module, and the embedding spaces alignment (ESA) module. The overall architecture of KSER is illustrated in Fig. \ref{Framework}. In this work, the embedding selection filter network (ESFNet) is designed as the KSF module, which effectively filters out the homogeneous and redundant information and retains the most relevant personalized knowledge for recommendation tasks. An attention-based mechanism is used to align the embedding spaces between semantic embeddings and feature embeddings. Note that there is no unique architectural design for the KSF and ESA modules.
\subsection{Knowledge Selection and Filtering}
Inspired by EPNet proposed to performs personalized selection on embedding for different users in multiple domains \cite{pepnet2023}, here, we proposed ESFNet to achieve knowledge selection and filter. In this architecture, differing from EPNet, the corresponding knowledge representation $\textbf{k}_i$ of each sample is first partitioned into $C$ vector chunks. Without loss of generality, it is assumed that the knowledge representation dimension $d_k$ is divisible by the number of chunks $C$. The chunk size is $s=d_k/C$. The knowledge representation $\textbf{k}_i$ can be rewritten as
\begin{equation}
    \label{Eq01}
\textbf{k}_i=\left[
\begin{array}{cccc}
\boldsymbol{k}_{i,[1:s]}^{(1)} &\dotsc & \boldsymbol{k}_{i,[1:s]}^{(L)} \\ 
\vdots & \vdots & \vdots \\ 
\boldsymbol{k}_{i,[(C-1)s+1:Cs]}^{(1)} &\dotsc & \boldsymbol{k}_{i,[(C-1)s+1:Cs]}^{(L)} \\ 
\end{array}
\right],
\end{equation}
where $\boldsymbol{k}_{i,[1:s]}^{(j)}\in\mathbb{R}^s$ denotes the knowledge embedding chunk of the $i$-th samples in $j$-th knowledge field. Then, adaptive weights are assigned to the different embedding chunks of different knowledge fields by integrating the feature fields of traditional recommenders, enabling the selection and filtering of knowledge embeddings. The weight matrix $\textbf{w}_i\in\mathbb{R}^{C\times{L}}$ is learned through a gate neural unit mentioned in \cite{pepnet2023}, which is formulated as
\begin{align}
    \label{Eq02}
    \textbf{w}_i=\;&\kappa\times\text{sigmod}(\text{Relu}(\textbf{z}_iW_1+b_1)W_2+b_2),\nonumber\\
    \textbf{z}_i=\;&\text{vec}^\mathsf{T}(\textbf{k}_i)\oplus(\oslash(\textbf{x}_i)),
\end{align}
where $W_1, W_2, b_1, b_2$ are the weights and biases of gate neural unit, $\kappa$ is the scaling factor, $\oplus$ denotes concatenation, $\text{vec}(\cdot)$ is the column-wise flattening operation of a matrix, $\oslash(\textbf{x}_i)$ operation means that the gradient backpropagation of $\textbf{x}_i$ is not performed. The selected knowledge embedding is computed by 
\begin{equation}
    \label{Eq03}
\bar{\textbf{k}}_i=\textbf{k}_i\otimes\textbf{w}_i,
\end{equation}
where $\otimes$ is the vector-wise product, $\textbf{k}_i,\bar{\textbf{k}}_i\in\mathbb{R}^{d_k\times{L}}$. All elements of each embedding chunk in different knowledge fields share an identical weight. ESFNet enables knowledge embedding selection and filtering without affecting the gradient updates of the recommendation backbone.
\subsection{Embedding Spaces Alignment}
After the knowledge selection and filtering, it is necessary to achieve the space alignment between the semantic embedding space and the feature space used to train the recommendation models. Then, we design an attention-based construction to achieve this goal.

For each knowledge field $\bar{\boldsymbol{k}}_{i}^{(j)}$ in selected knowledge embeddings $\bar{\textbf{k}}_i$, the cross attention is employed to extract the most relevant information from $\bar{\boldsymbol{k}}_{i}^{(j)}$ with respect to the feature space of the recommendation model. Before performing cross attention, the following processing is first applied to the knowledge embeddings $\bar{\boldsymbol{k}}_{i}^{(j)}$ for each knowledge field.
\begin{equation}
    \label{Eq04}
    \breve{\boldsymbol{k}}_{i}^{(j)}=\text{Dense}(\text{Relu}(\text{Dense}(\bar{\boldsymbol{k}}_{i}^{(j)}))).
\end{equation}
Then the processed knowledge vector $\breve{\boldsymbol{k}}_{i}^{(j)}$ and the embedding vector $E(\textbf{x}_i)\in\mathbb{R}^{d_e}$ of traditional feature fields are partitioned into $C_k$ and $C_x$ chunks, and then stacked into a knowledge matrix $\tilde{\textbf{k}}_i^{(j)}\in\mathbb{R}^{C_k\times{d_k}}$ in the $j$-th knowledge field and the traditional feature embedding matrix $\tilde{\textbf{x}}_i\in\mathbb{R}^{C_x\times{d_x}}$, respectively. 

Then, the cross attention is performed on $\tilde{\textbf{k}}_i^{(j)}$ and $\tilde{\textbf{x}}_i$, which is formulated as 
\begin{align}
    \label{Eq05}   
\textbf{Q},\textbf{K},\textbf{V}=\;&\tilde{\textbf{x}}_iW_Q,\tilde{\textbf{k}}_i^{(j)}W_K,\tilde{\textbf{k}}_i^{(j)}W_V,\nonumber\\
    \textbf{o}_i^{(j)}=\;&\text{softmax}(\textbf{Q}\textbf{K}^{\mathsf{T}})\textbf{V},
\end{align}
where $\textbf{o}_i^{(j)}\in\mathbb{R}^{C_x\times{m}}$, $W_Q\in\mathbb{R}^{d_x\times{n}}$, $W_K\in\mathbb{R}^{d_k\times{n}}$, $W_V\in\mathbb{R}^{d_k\times{m}}$ are learnable weight matrices. Finally, a self-attention is used to perform the knowledge fusion across fields.
\begin{equation}
    \label{Eq06}
    \textbf{o}_i=\text{SelfAttn}(\textbf{o}_i^{(1)}\oplus\textbf{o}_i^{(2)}\oplus\dotsc\oplus\textbf{o}_i^{(L)})
\end{equation}
The ESFNet and ESA pseudocodes are given in Section Details of ESFNet and ESA of Supplementary Material.
\subsection{KSER Training Strategies}
To seamlessly integrate the outputs of the ESA module into the recommendation models, we propose two training strategies, namely all-parameters training and extractor-only training strategies, which can be selected according to different application scenarios.

\subsubsection{All-Parameters Training}
Once the output of the ESA module is obtained, the input of recommendation models with all-parameters training strategy is formulated as
\begin{equation}
    \label{Eq07}
    \textbf{e}_i=\text{vec}(\textbf{o}_i)\oplus{E}(\textbf{x}_i).
\end{equation}
The processed knowledge embedding vector $\textbf{o}_i$ can be considered as additional feature fields of the traditional recommender systems. The prediction value is represented as
\begin{equation}
    \label{Eq08}
    \hat{y}=\text{M}(\textbf{e}_i),
\end{equation}
where $\text{M}(\cdot)$ denotes the traditional recommendation models. For different recommendation tasks, different loss function can be used to train models. In the training process, the world knowledge filtering network, embedding spaces alignment module, and the recommendation backbone model are jointly optimized.
\begin{table*}[t]
\centering
\setlength{\tabcolsep}{1mm}
\fontsize{9}{\baselineskip}\selectfont
\begin{tabular}{l *{14}{c}}
\toprule
 & \multicolumn{6}{c}{MovieLens-1M} & \multicolumn{6}{c}{Amazon-Book} \\
\cmidrule(lr){2-7} \cmidrule(lr){8-13}
Backbone
& \multicolumn{3}{c}{AUC $\uparrow$} & \multicolumn{3}{c}{LogLoss $\downarrow$} 
& \multicolumn{3}{c}{AUC $\uparrow$} & \multicolumn{3}{c}{LogLoss $\downarrow$} \\
\cmidrule(lr){2-4} \cmidrule(lr){5-7} \cmidrule(lr){8-10} \cmidrule(lr){11-13}
  & KAR & KSER & Improv. 
                & KAR & KSER & Improv. 
                & KAR & KSER & Improv. 
                & KAR & KSER & Improv. \\
\midrule
% 示例数据行（你可以替换为实际数值）
DIN &0.78779  &\textbf{0.78803} & 0.0305\% 
        &0.54881  &\textbf{0.54607}  & 0.4993\% 
         &0.83349  &\textbf{0.83785}  & 0.5231\% 
        &0.48898  &\textbf{0.48325}  & 1.1718\% \\
DCNv1 &\textbf{0.7871}  &0.78673  & -0.0470\% 
        &0.54964  &\textbf{0.54734}  &0.4185 \% 
        &0.83153  &\textbf{0.83296}  &0.1720 \% 
        &0.49276  &\textbf{0.48892}  &0.7793 \% \\
DCNv2 &\textbf{0.78765}  &0.78654  & -0.1409\% 
        &0.54911  &\textbf{0.54774}  &0.2495 \% 
        &0.83164  &\textbf{0.83331}  &0.2008 \% 
        &0.49224  &\textbf{0.48862}  &0.7354 \% \\
DIEN &0.78814  &\textbf{0.78922}  &0.1370 \% 
        &0.54767  &\textbf{0.54356}  &0.7505 \% 
        &0.83286  &\textbf{0.83621}  &0.4022\% 
        &0.49079  &\textbf{0.48499}  &1.1818 \% \\
DeepFM &0.78582  &\textbf{0.78584}  &0.0025\% 
        &0.55018  &\textbf{0.54879}  &0.2526 \% 
        &0.83183  &\textbf{0.83329}  &0.1755\% 
        &0.49314  &\textbf{0.48445}  &1.7622 \% \\
FiBiNet &0.78798  &\textbf{0.78815}  &0.0216 \% 
        &0.54826  &\textbf{0.54769}  &0.1040 \% 
        &0.83203  &\textbf{0.83311}  &0.1298 \% 
        &0.49553  &\textbf{0.48918}  &1.2815 \% \\
AutoInt &\textbf{0.78675}  &0.7864  &-0.0445 \% 
        &0.54914  &\textbf{0.54761}  &0.2786 \% 
        &0.83231  &\textbf{0.83301}  &0.0841 \% 
        &0.49208  &\textbf{0.48959}  &0.5060 \% \\
FiGNN &0.78757  &\textbf{0.78814}  & 0.0724\% 
        &0.54986  &\textbf{0.54832}  &0.2801 \% 
        &0.83217  &\textbf{0.83322}  &0.1262 \% 
        &0.49285  &\textbf{0.48888}  &0.8055 \% \\
xDeepFM &0.78794  &\textbf{0.78884}  &0.1142 \% 
        &0.54805  &\textbf{0.54491}  &0.5729 \% 
        &0.83221  &\textbf{0.83303}  &0.0985 \% 
        &0.49245  &\textbf{0.48943}  &0.6133 \% \\
\bottomrule
\end{tabular}
\caption{Performances of the \textbf{all-parameters training} strategy and baselines.}
\label{Tab_0401}
\end{table*}
\subsubsection{Extractor-Only Training}
To guarantee the performance stability of the online recommender and further reduce the computational cost, the extractor-only training strategy can be used to achieve this objective with minimal modifications to the backbone model. As shown in Fig. \ref{Framework}, the selected knowledge embeddings have explicitly interacted with the feature of the traditional recommendation models in ESA module. Therefore, it is not necessary to jointly train the world knowledge filtering and embedding spaces alignment modules with the backbone model if the backbone model has been fully trained. Then, the way of integrating knowledge embeddings into the model is formulated as 
\begin{equation}
    \label{Eq09}
    \hat{y}=\text{ACT}(\text{Dense}(\text{vec}(\textbf{o}_i)\oplus\text{M}'(E(\textbf{x}_i)))),
\end{equation}
where $\text{ACT}(\cdot)$ is the activation function, $\text{M}'(\cdot)$ is the backbone recommendation model without the embedding layer and the output layer. During training, only world knowledge filtering and embedding spaces alignment modules, collectively referred to as the extractor, along with the embedding layer and output layer of the backbone model perform the parameter updates. The parameters of $\text{M}'(\cdot)$ are frozen. Compared with the entire extraction architecture, the number of the parameters of the embedding and output layers can be neglected. With this operation, the extractor-only training strategy not only improves model performance with minor changes but also ensures better compatibility with the online-serving recommendation models.

\section{Experiments}
In this section, the extensive experiments are performed to compare the performance of the present KSER framework with existing SOTA approaches and to answer the following questions: 
\begin{itemize}
\item Q1: Does the developed KSER framework with all-parameters training strategy improve the prediction performance compared with the SOTA? 
\item Q2: Does the two training strategy enhance the performance of the traditional recommenders? 
\item Q3: What are the effects of the two core modules in the developed framework? 
\item Q4: Does KSER achieve effective selection and exploitation of high-quality knowledge from LLMs for recommendation?
\end{itemize}
\subsection{Experimental Setup}
\subsubsection{Dataset and Preprocessing.} Two public datasets are used to conduct the experiments, namely MovieLens-1M with 1 million ratings of 6000 users to 4000 movies and Amazon-Book \cite{JustifyRec}. Similar to literature \cite{DIN,TowardsOpen-WorldRec}, the rating scores in MovieLens-1M are binarized as 1 when it is greater than four, otherwise as 0. For the Amazon-Book dataset, the rating scores are binarized as 0 when it is less than 5, otherwise as 1. The prompt designs, knowledge generation, and knowledge encoding is consistent with the literature \cite{TowardsOpen-WorldRec}. The reasoning knowledge on user preferences and the factual knowledge on items from LLMs \cite{TowardsOpen-WorldRec} are used to construct the knowledge repository $\textbf{K}$.
\subsubsection{Baselines and Experimental Details.}
The representative approaches--\textbf{DIN} \cite{DIN}, \textbf{DCNv1} \cite{DCN}, \textbf{DCNv2} \cite{DCNv2}, \textbf{DIEN} \cite{DIEN}, \textbf{DeepFM} \cite{DeepFM}, \textbf{FiBiNet} \cite{FiBiNet}, \textbf{AutoInt} \cite{AutoInt}, \textbf{FiGNN} \cite{FiGNN}, \textbf{xDeepFM} \cite{xDeepFM}-- are employed as backbone models. We compare the KESR framework with these backbone methods and the SOTA method, i.e., knowledge augmented recommendation (KAR) \cite{TowardsOpen-WorldRec}. We use NVIDIA A100-SXM4-80GB GPU to train and test all models on all datasets.

Here, we adopt the two common metrics used in recommender system, i.e., area under the ROC curve (AUC), and binary cross-entropy loss (LogLoss), to evaluate the performance of prediction models. To guarantee the fair comparison, the best performance for every backbone and sota model is recorded by tuning the learning rates in $\{1e-4, 5e-4,1e-3\}$. Besides, the numbers of the preference experts and item experts in the KAR scheme are tuned in $\{2, 3, 4, 5, 6\}$. For the extractor-only training strategy, we first train the backbone model. Then, the parameters of backbone models without the embedding and output layers are frozen. Finally, we integrate the processed knowledge embeddings into the backbone models according to (\ref{Eq09}) and update the parameters of the extractor as well as the embedding and output layers of the backbone model.
\subsection{Performance Comparison (for Q1 and Q2)}
\begin{figure}[t]
\centering
\includegraphics[width=0.6\columnwidth]{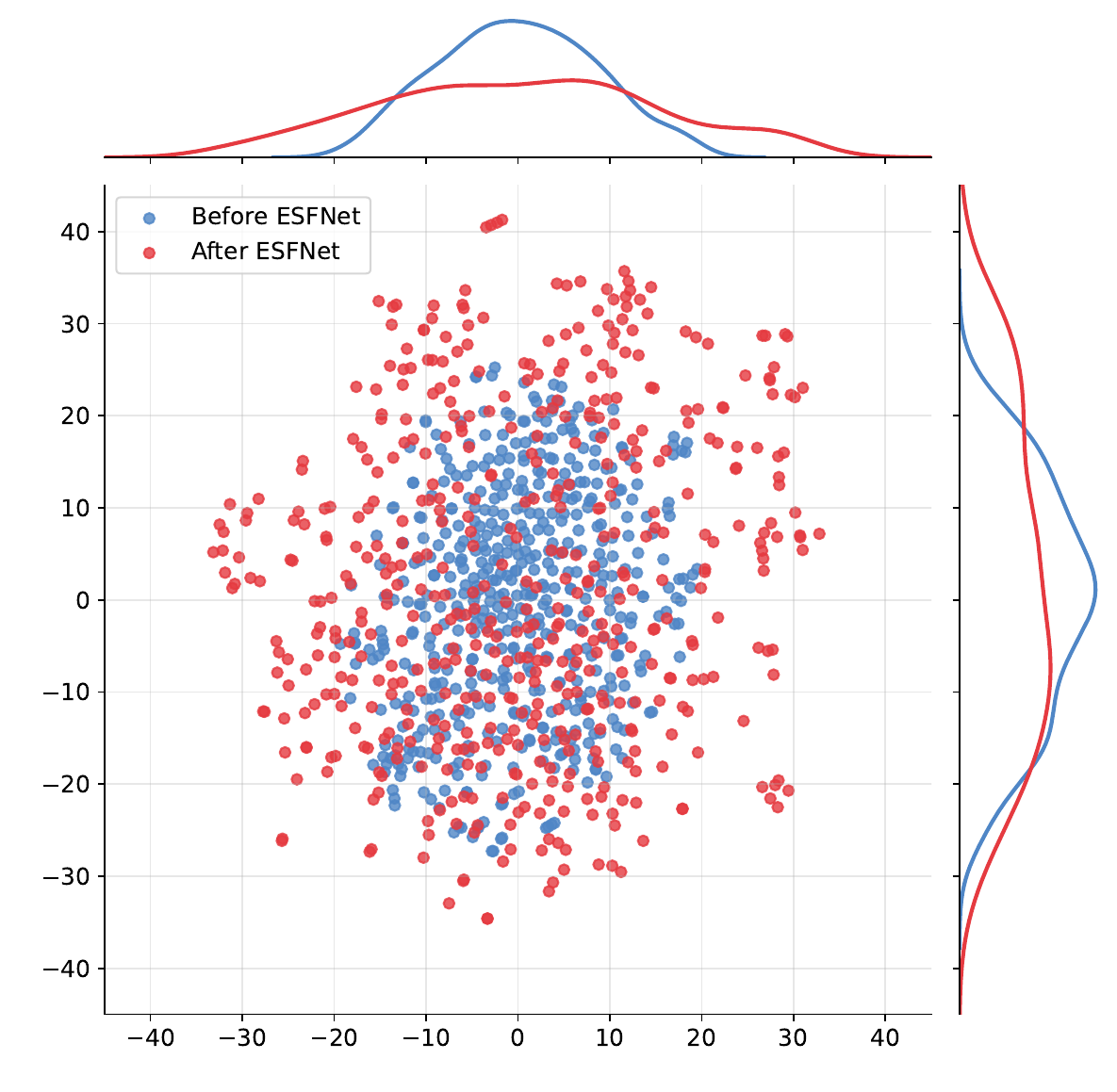} 
\caption{Embedding vectors before and after passing through ESFNet using T-SNE.}
\label{esfnet}
\end{figure}
\begin{figure}[t]
\centering
\includegraphics[width=1\columnwidth]{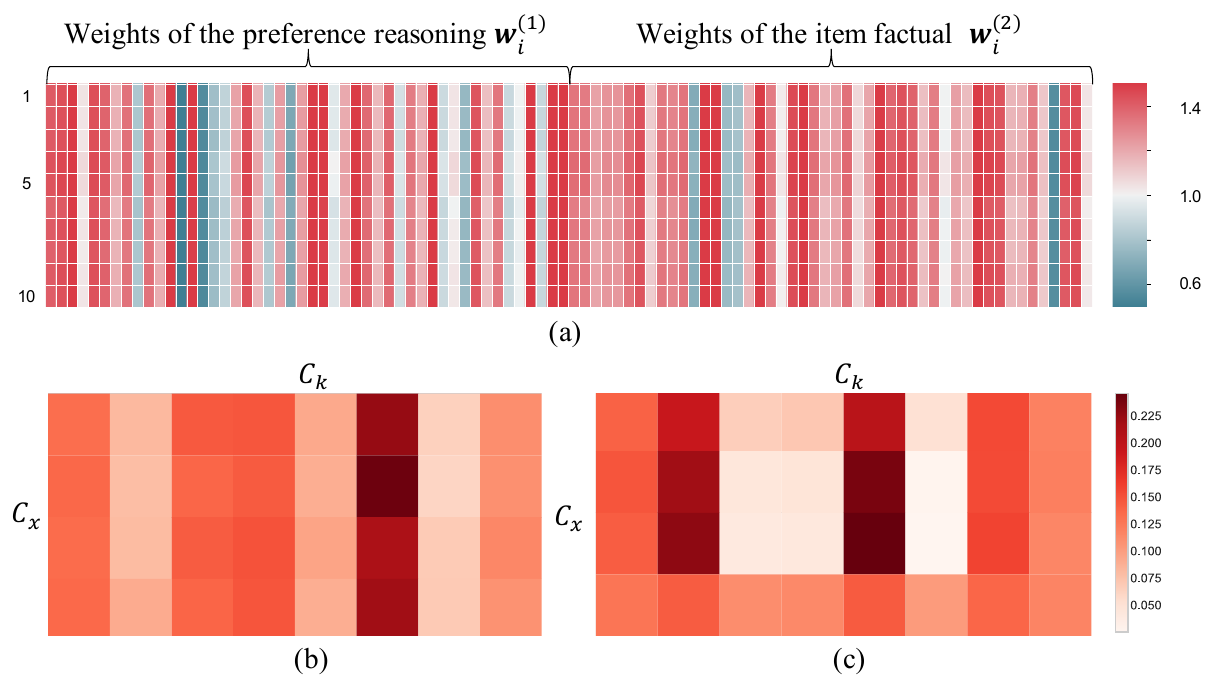} 
\caption{(a) The weights from the gate neural unit for preference reasoning and item factual; (b) Attention score of the user preference filed in cross attention; (c) Attention score of the item factual filed in cross attention.}
\label{attscoreesfnetweight}
\end{figure}

\begin{table*}[t]
\centering
\setlength{\tabcolsep}{1mm}
\fontsize{9}{\baselineskip}\selectfont
\begin{tabular}{l *{12}{c}}
\toprule
 & \multicolumn{6}{c}{MovieLens-1M} & \multicolumn{6}{c}{Amazon-Book} \\
\cmidrule(lr){2-7} \cmidrule(lr){8-13}
Backbone 
& \multicolumn{3}{c}{AUC $\uparrow$} & \multicolumn{3}{c}{LogLoss $\downarrow$} 
& \multicolumn{3}{c}{AUC $\uparrow$} & \multicolumn{3}{c}{LogLoss $\downarrow$} \\
\cmidrule(lr){2-4} \cmidrule(lr){5-7} \cmidrule(lr){8-10} \cmidrule(lr){11-13}
 & Base & KSER & Improv. 
 & Base & KSER & Improv. 
 & Base & KSER & Improv. 
 & Base & KSER & Improv. \\
\midrule
% 示例数据行（你可以替换为实际数值）
DIN     &0.77269 &\textbf{0.78449} &1.5271\% 
        &0.56072 &\textbf{0.55554} &0.9238\% 
        &0.82337 &\textbf{0.83031} &0.8429\% 
        &0.50260 &\textbf{0.49526} &1.4604\% \\
DCNv1   &0.77222 &\textbf{0.78257} &1.3403\%	
        &0.56217 &\textbf{0.55161} &1.8784\% 
        &0.82578 &\textbf{0.83282} &0.8525\% 
        &0.49823 &\textbf{0.48965} &1.7221\% \\
DCNv2   &0.77238 &\textbf{0.78428} &1.5407\%	
        &0.56165 &\textbf{0.55074} &1.9425\%
        &0.82596 &\textbf{0.83270} &0.8160\% 
        &0.49963 &\textbf{0.48934} &2.0595\% \\
DIEN    &0.75709 &\textbf{0.78387} &3.5372\%	
        &0.57662 &\textbf{0.55567} &3.6332\%
        &0.80546 &\textbf{0.82942} &2.9747\% 
        &0.52403 &\textbf{0.49403} &5.7249\% \\
DeepFM  &0.77221 &\textbf{0.77901} &0.8806\%	
        &0.56155 &\textbf{0.55517} &1.1361\%
        &0.82601 &\textbf{0.83190} &0.7131\% 
        &0.49845 &\textbf{0.49150} &1.3943\% \\
FiBiNet &0.77214 &\textbf{0.78368} &1.4945\%	
        &0.56230 &\textbf{0.55088} &2.0309\%
        &0.82544 &\textbf{0.83235} &0.8371\% 
        &0.49878 &\textbf{0.49093} &1.5738\% \\
AutoInt &0.77260 &\textbf{0.78321} &1.3733\%	
        &0.56153 &\textbf{0.55213} &1.6740\%
        &0.82460 &\textbf{0.83113} &0.7919\% 
        &0.50002 &\textbf{0.49297} &1.4099\% \\
FiGNN   &0.77214 &\textbf{0.78295} &1.4000\%	
        &0.56184 &\textbf{0.55305} &1.5645\%
        &0.82583 &\textbf{0.83153} &0.6902\% 
        &0.49883 &\textbf{0.49384} &1.0003\% \\
xDeepFM &0.77254 &\textbf{0.78345} &1.4122\%	
        &0.56129 &\textbf{0.55215} &1.6284\%
        &0.82501 &\textbf{0.83193} &0.8388\% 
        &0.50016 &\textbf{0.49057} &1.9174\% \\
\bottomrule
\end{tabular}
\caption{Performances under the \textbf{extractor-only
training} strategy and backbone models.}
\label{Tab_0402}
\end{table*}
For the CTR prediction task, the performance of KAR and KSER with the all-parameters training strategy across various backbone recommendation models is presented in Table \ref{Tab_0401}, where the improvements in AUC and reductions in LogLoss are computed as 
$\text{improv.}=\frac{\text{AUC}_\text{KSER}-\text{AUC}_\text{KAR}}{\text{AUC}_\text{KSER}}$,
$\text{improv.}=\frac{\text{LogLoss}_\text{KAR}-\text{LogLoss}_\text{KSER}}{\text{LogLoss}_\text{KAR}}$,
% \begin{align}
% \text{improv.}=&\;\frac{\text{AUC}_\text{KSER}-\text{AUC}_\text{KAR}}{\text{AUC}_\text{KSER}},\nonumber\\
% \text{improv.}=&\;\frac{\text{LogLoss}_\text{KAR}-\text{LogLoss}_\text{KSER}}{\text{LogLoss}_\text{KAR}},
% \end{align}
respectively. For almost all of metrics and datasets, the performances of KSER with the all-parameters training strategy outperform the KAR method, which implies that the knowledge filtering and embedding spaces alignment modules can select the preferences and factual knowledge beneficial to the recommendation model and achieve feature embedding fusion across different spaces. Compared the experimental results of KSER with KAR, it is crucial for enhancing the performance of recommendation models to accurately select and sufficiently leverage knowledge from LLMs.

For the extractor-only training strategy, the experimental results are given in Table \ref{Tab_0402}. It can be observed that KSER consistently outperforms the backbone models, achieving significant improvements in AUC and reductions in LogLoss across all datasets. The extractor-only training strategy opens a new direction for the knowledge-enhanced recommendation based on LLMs. It enables the significant improvement of the backbone recommender performance solely through the efficient fusion and utilization of world and reasoning knowledge without retraining the backbone models. 

Comparing the all-parameters training strategy with the extractor-only strategy as shown in Tables \ref{Tab_0401} and \ref{Tab_0402}, we observe that the former achieves superior performance on all datasets. According to Eqs. (\ref{Eq07})-(\ref{Eq09}), it can be observed that, for different training strategies, the concatenating positions and interaction mechanisms between knowledge vectors $\textbf{o}_i$ and feature embeddings ${E}(\textbf{x}_i)$ are different. The all-parameters training strategy can further leverage the structural advantages of backbone models for simultaneous processing of both knowledge vectors and feature embeddings. 

\subsection{Ablation Studies (for Q3 and Q4)}
\begin{table*}[t]
\centering
\setlength{\tabcolsep}{1mm}
\fontsize{9}{\baselineskip}\selectfont
\begin{tabular}{l *{12}{c}}
\toprule
 & \multicolumn{6}{c}{MovieLens-1M} & \multicolumn{6}{c}{Amazon-Book} \\
\cmidrule(lr){2-7} \cmidrule(lr){8-13}
Backbone
& \multicolumn{3}{c}{AUC $\uparrow$} & \multicolumn{3}{c}{LogLoss $\downarrow$} 
& \multicolumn{3}{c}{AUC $\uparrow$} & \multicolumn{3}{c}{LogLoss $\downarrow$} \\
\cmidrule(lr){2-4} \cmidrule(lr){5-7} \cmidrule(lr){8-10} \cmidrule(lr){11-13}
&{\makecell[c]{KSER \\ w/o \\ ESFNet}} & {\makecell[c]{KSER \\ w/o \\ ESA}}  &KSER 
&{\makecell[c]{KSER \\ w/o \\ ESFNet}} & {\makecell[c]{KSER \\ w/o \\ ESA}}  &KSER 
&{\makecell[c]{KSER \\ w/o \\ ESFNet}} & {\makecell[c]{KSER \\ w/o \\ ESA}} &KSER  
&{\makecell[c]{KSER \\ w/o \\ ESFNet}} & {\makecell[c]{KSER \\ w/o \\ ESA}}  &KSER \\
\midrule
DIN &0.78733  &0.78591  &\textbf{0.78803}  
        &0.54656  &0.54815  &\textbf{0.54607}  
        &0.83759  &0.82327  &\textbf{0.83785}  
        &0.48416  &0.50244  &\textbf{0.48325}   \\
DCNv1 &0.78584  &\textbf{0.78741}  &0.78673   
        &0.54818  &0.54754  &\textbf{0.54734} 
        &0.83278  &0.83253  &\textbf{0.83296} 
        &0.48921  &0.48958  &\textbf{0.48892}  \\
DCNv2 &0.7859  &\textbf{0.78849}  &0.78654   
        &0.54833  &0.54815  &\textbf{0.54774}  
        &0.83316  &0.83266  &\textbf{0.83331}  
        &0.48878  &0.48933  &\textbf{0.48862}   \\
DIEN &0.78756  &0.78912  &\textbf{0.78922}  
        &0.54749  &0.54506  &\textbf{0.54356}  
        &0.83598  &0.80504  &\textbf{0.83621}  
        &0.48515  &0.52873  &\textbf{0.48499}   \\
DeepFM &0.78447  &0.78538  &\textbf{0.78584} 
        &0.54989  &0.54954  &\textbf{0.54879}  
        &0.83264  &0.83261  &\textbf{0.83329}  
        &0.48954  &0.49139  &\textbf{0.48445}   \\
FiBiNet &0.78637  &0.78724  &\textbf{0.78815}  
        &0.55036  &0.54801  &\textbf{0.54769}   
        &0.83277  &0.8327  &\textbf{0.83311}   
        &0.48968  &0.49037  &\textbf{0.48918}   \\
AutoInt &0.78584  &\textbf{0.78718}  &0.7864   
        &0.5482  &0.55022  &\textbf{0.54761}   
        &0.83266  &0.83114  &\textbf{0.83301}   
        &0.48987  &0.49201  &\textbf{0.48959}   \\
FiGNN &0.7869  &0.787  &\textbf{0.78814}   
        &0.54853  &0.54877  &\textbf{0.54832}   
        &0.83298  &0.8322  &\textbf{0.83322}   
        &0.48913  &0.49123  &\textbf{0.48888}   \\
xDeepFM &0.78584  &0.54877  &\textbf{0.78884}   
        &0.54925  &0.54604  &\textbf{0.54491}   
        &0.83292  &0.83226  &\textbf{0.83303}   
        &0.48973  &0.4919  &\textbf{0.48943}   \\
\bottomrule
\end{tabular}
\caption{Results of the ablation study for KSER with the all-parameters training strategy.}
\label{Tab_0403}
\end{table*}
\begin{figure*}[t]
\centering
\includegraphics[width=1\textwidth]{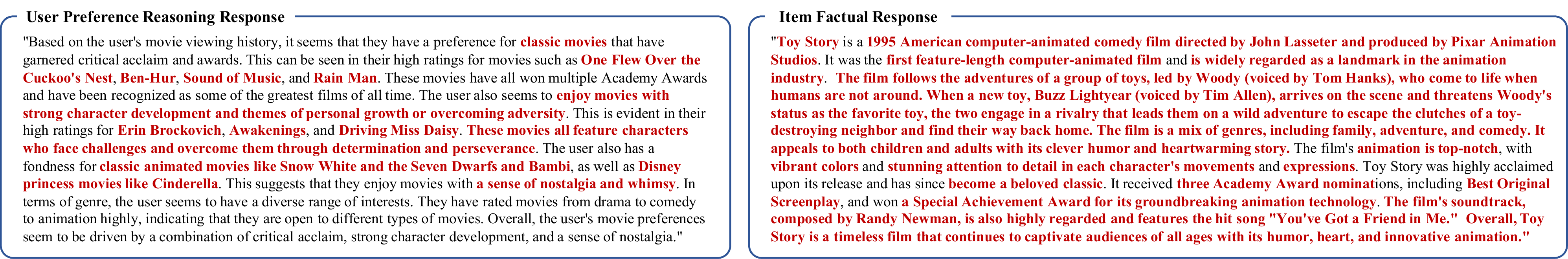} 
\caption{(a) User preference reasoning response; (b) Item factual response.}
\label{casestudy}
\end{figure*}
\subsubsection{Effects of ESFNet And ESA Modules}
Under the all-parameters training strategy, we conduct ablation experiments on different submodules of KSER to analyze the effects of ESFNet and ESA Modules. The specific model structures of the KSER variants are illustrated as follows:
\begin{itemize}
    \item KSER (w/o ESFNet): ESFNet is directly removed from KSER.
    \item KSER (w/o ESA): ESA is replaced with a dense projection layer to align the dimensions of the knowledge embedding and the traditional feature embedding.
\end{itemize}
The ablation results are presented in Table~\ref{Tab_0403}. Compared with the complete KSER framework, KSER (w/o ESFNet) exhibits performance degradation, demonstrating that ESFNet effectively selects beneficial knowledge and filters out redundant information from LLMs. To further validate the effectiveness of the ESFNet, we randomly select 512 samples and visualize their embedding vectors (before and after passing through ESFNet) in Fig. \ref{esfnet} using T-SNE for dimensionality reduction, where red and blue lines represent distributions of embedding vectors before and after passing through ESFNet along the two dimensions, respectively. It can be observed that the knowledge embeddings after passing through ESFNet become more discriminative and the distribution of these embeddings becomes smoother since ESFNet selects high-quality, relevant knowledge for each sample, filters out redundant information, and thereby helps the recommendation system reduce its modeling burden. On the other hand, according to Table \ref{Tab_0403}, the performance of KSER (w/o ESA) also decline across almost all of metrics and datasets. Therefore, it is necessary to align the embedding spaces between the semantic embedding vectors and the feature vectors used in the traditional recommendation models.

Next, 10 samples are randomly selected to observe the knowledge selection and filtering.
Their weights from the gate neural unit for the reasoning knowledge on user preferences and factual knowledge on items are given in Fig. \ref{attscoreesfnetweight}(a). 
According to the weight distribution, we can conclude that the proportion of high-quality knowledge in the preference reasoning knowledge field is higher than that in item factual knowledge field. Fig. \ref{casestudy} provides two examples of the user preference reasoning response and item factual response from LLMs that received different weight distributions, where the red words are the beneficial text and others can be regarded as redundant knowledge. The proportion of high-quality knowledge in the preference reasoning and item
factual responses are different, which is consistent with the experimental results. The attention scores between $E(\textbf{x}_i)$ chunks and $\breve{\boldsymbol{k}}_{i}^{(j)}$ chunks of different fields in ESA module are shown in Fig. \ref{attscoreesfnetweight}(b) and (c), which implies that the relevant information from $\bar{\boldsymbol{k}}_{i}^{(j)}$ with respect to $E(\textbf{x}_i)$ can be effectively extracted.

% \subsubsection{Effects of The Number of Chunks in ESFNet}
\section{Conclusions}
This work focuses on the effective selection and exploitation of high-quality knowledge from LLMs, and proposes the KSER framework with two training strategies to be compatible with a broader range of application scenarios. The problem of the homogeneous, redundant, and even erroneous information in LLMs responses can be effectively addressed by using the designed ESFNet. Besides, the attention-based architecture can be used to achieve embedding spaces alignment. Remarkably, the proposed KSER framework can achieve the performance improvement of backbone models solely through updating the parameters of the ESFNet and ESA module, which provides a novel perspective on knowledge-augmented recommendation based on LLMs. In the extractor-only training strategy, knowledge-augmented recommendation is further extended to a new paradigm, namely the selection \& alignment-only training paradigm. The future works can focus on the selection \& alignment modules designs, which can be trained based on a trained discriminative recommendation model. Experimental results verify the performance of the KSER framework, the effectiveness of the designed modules, and the efficiency of the training strategies.
\bibliography{aaai2026}

\begin{thebibliography}{29}
\providecommand{\natexlab}[1]{#1}

\bibitem[{Chang et~al.(2023)Chang, Zhang, Hui, Leng, Niu, Song, and Gai}]{pepnet2023}
Chang, J.; Zhang, C.; Hui, Y.; Leng, D.; Niu, Y.; Song, Y.; and Gai, K. 2023.
\newblock PEPNet: Parameter and Embedding Personalized Network for Infusing with Personalized Prior Information.
\newblock In \emph{Proceedings of the 29th ACM SIGKDD Conference on Knowledge Discovery and Data Mining}, KDD '23, 3795–3804. New York, NY, USA: Association for Computing Machinery.
\newblock ISBN 9798400701030.

\bibitem[{Deng et~al.(2025)Deng, Wang, Cai, Ren, Hu, Ding, Luo, and Zhou}]{OneRec}
Deng, J.; Wang, S.; Cai, K.; Ren, L.; Hu, Q.; Ding, W.; Luo, Q.; and Zhou, G. 2025.
\newblock OneRec: Unifying Retrieve and Rank with Generative Recommender and Iterative Preference Alignment.
\newblock arXiv:2502.18965.

\bibitem[{Devlin et~al.(2019)Devlin, Chang, Lee, and Toutanova}]{BERT}
Devlin, J.; Chang, M.-W.; Lee, K.; and Toutanova, K. 2019.
\newblock {BERT}: Pre-training of Deep Bidirectional Transformers for Language Understanding.
\newblock In Burstein, J.; Doran, C.; and Solorio, T., eds., \emph{Proceedings of the 2019 Conference of the North {A}merican Chapter of the Association for Computational Linguistics: Human Language Technologies, Volume 1 (Long and Short Papers)}, 4171--4186. Minneapolis, Minnesota: Association for Computational Linguistics.

\bibitem[{Gu et~al.(2025)Gu, Zhong, Xia, Yang, Lu, Jiang, and Gai}]{R4ec}
Gu, H.; Zhong, R.; Xia, Y.; Yang, W.; Lu, C.; Jiang, P.; and Gai, K. 2025.
\newblock R4ec: A Reasoning, Reflection, and Refinement Framework for Recommendation Systems.
\newblock arXiv:2507.17249.

\bibitem[{Guo et~al.(2017)Guo, TANG, Ye, Li, and He}]{DeepFM}
Guo, H.; TANG, R.; Ye, Y.; Li, Z.; and He, X. 2017.
\newblock DeepFM: A Factorization-Machine based Neural Network for CTR Prediction.
\newblock In \emph{Proceedings of the Twenty-Sixth International Joint Conference on Artificial Intelligence, {IJCAI-17}}, 1725--1731.

\bibitem[{Han et~al.(2025)Han, Yin, Chen, Jiang, Jiang, Li, Ma, Huang, Li, Jing, Han, Zhou, Yu, Liu, and Lin}]{MTGR}
Han, R.; Yin, B.; Chen, S.; Jiang, H.; Jiang, F.; Li, X.; Ma, C.; Huang, M.; Li, X.; Jing, C.; Han, Y.; Zhou, M.; Yu, L.; Liu, C.; and Lin, W. 2025.
\newblock MTGR: Industrial-Scale Generative Recommendation Framework in Meituan.
\newblock arXiv:2505.18654.

\bibitem[{Hu et~al.(2025)Hu, Li, Jiao, Nakagawa, Deng, Cai, Zhou, and Ren}]{BridgUser}
Hu, Z.; Li, Z.; Jiao, Z.; Nakagawa, S.; Deng, J.; Cai, S.; Zhou, T.; and Ren, F. 2025.
\newblock Bridging the User-side Knowledge Gap in Knowledge-aware Recommendations with Large Language Models.
\newblock \emph{Proceedings of the AAAI Conference on Artificial Intelligence}, 39(11): 11799--11807.

\bibitem[{Huang, Zhang, and Zhang(2019)}]{FiBiNet}
Huang, T.; Zhang, Z.; and Zhang, J. 2019.
\newblock FiBiNET: combining feature importance and bilinear feature interaction for click-through rate prediction.
\newblock In \emph{Proceedings of the 13th ACM Conference on Recommender Systems}, RecSys '19, 169–177. New York, NY, USA: Association for Computing Machinery.
\newblock ISBN 9781450362436.

\bibitem[{Huang et~al.(2025)Huang, Lian, Lei, Yao, Lian, and Xie}]{RecAIAgent}
Huang, X.; Lian, J.; Lei, Y.; Yao, J.; Lian, D.; and Xie, X. 2025.
\newblock Recommender AI Agent: Integrating Large Language Models for Interactive Recommendations.
\newblock \emph{ACM Trans. Inf. Syst.}, 43(4).

\bibitem[{Kang and McAuley(2018)}]{SASRec}
Kang, W.-C.; and McAuley, J. 2018.
\newblock Self-Attentive Sequential Recommendation.
\newblock In \emph{2018 IEEE International Conference on Data Mining (ICDM)}, 197--206.

\bibitem[{Li et~al.(2019)Li, Cui, Wu, Zhang, and Wang}]{FiGNN}
Li, Z.; Cui, Z.; Wu, S.; Zhang, X.; and Wang, L. 2019.
\newblock Fi-GNN: Modeling Feature Interactions via Graph Neural Networks for CTR Prediction.
\newblock In \emph{Proceedings of the 28th ACM International Conference on Information and Knowledge Management}, CIKM '19, 539–548. New York, NY, USA: Association for Computing Machinery.
\newblock ISBN 9781450369763.

\bibitem[{Lian et~al.(2018)Lian, Zhou, Zhang, Chen, Xie, and Sun}]{xDeepFM}
Lian, J.; Zhou, X.; Zhang, F.; Chen, Z.; Xie, X.; and Sun, G. 2018.
\newblock xDeepFM: Combining Explicit and Implicit Feature Interactions for Recommender Systems.
\newblock In \emph{Proceedings of the 24th ACM SIGKDD International Conference on Knowledge Discovery \& Data Mining}, KDD '18, 1754–1763. New York, NY, USA: Association for Computing Machinery.
\newblock ISBN 9781450355520.

\bibitem[{Ni, Li, and McAuley(2019)}]{JustifyRec}
Ni, J.; Li, J.; and McAuley, J. 2019.
\newblock Justifying Recommendations using Distantly-Labeled Reviews and Fine-Grained Aspects.
\newblock In Inui, K.; Jiang, J.; Ng, V.; and Wan, X., eds., \emph{Proceedings of the 2019 Conference on Empirical Methods in Natural Language Processing and the 9th International Joint Conference on Natural Language Processing (EMNLP-IJCNLP)}, 188--197. Hong Kong, China: Association for Computational Linguistics.

\bibitem[{Radford et~al.(2018)Radford, Narasimhan, Salimans, Sutskever et~al.}]{radford2018improving}
Radford, A.; Narasimhan, K.; Salimans, T.; Sutskever, I.; et~al. 2018.
\newblock Improving language understanding by generative pre-training.

\bibitem[{Rajput et~al.(2023)Rajput, Mehta, Singh, Keshavan, Vu, Heidt, Hong, Tay, Tran, Samost, Kula, Chi, and Sathiamoorthy}]{Tiger}
Rajput, S.; Mehta, N.; Singh, A.; Keshavan, R.; Vu, T.; Heidt, L.; Hong, L.; Tay, Y.; Tran, V.~Q.; Samost, J.; Kula, M.; Chi, E.~H.; and Sathiamoorthy, M. 2023.
\newblock Recommender systems with generative retrieval.
\newblock In \emph{Proceedings of the 37th International Conference on Neural Information Processing Systems}, NIPS '23. Red Hook, NY, USA: Curran Associates Inc.

\bibitem[{Song et~al.(2019)Song, Shi, Xiao, Duan, Xu, Zhang, and Tang}]{AutoInt}
Song, W.; Shi, C.; Xiao, Z.; Duan, Z.; Xu, Y.; Zhang, M.; and Tang, J. 2019.
\newblock AutoInt: Automatic Feature Interaction Learning via Self-Attentive Neural Networks.
\newblock In \emph{Proceedings of the 28th ACM International Conference on Information and Knowledge Management}, CIKM '19, 1161–1170. New York, NY, USA: Association for Computing Machinery.
\newblock ISBN 9781450369763.

\bibitem[{Vaswani et~al.(2017)Vaswani, Shazeer, Parmar, Uszkoreit, Jones, Gomez, Kaiser, and Polosukhin}]{NIPS2017_3f5ee243}
Vaswani, A.; Shazeer, N.; Parmar, N.; Uszkoreit, J.; Jones, L.; Gomez, A.~N.; Kaiser, L.~u.; and Polosukhin, I. 2017.
\newblock Attention is All you Need.
\newblock In Guyon, I.; Luxburg, U.~V.; Bengio, S.; Wallach, H.; Fergus, R.; Vishwanathan, S.; and Garnett, R., eds., \emph{Advances in Neural Information Processing Systems}, volume~30. Curran Associates, Inc.

\bibitem[{Wang et~al.(2025)Wang, Zhang, Yang, Chen, Tang, Zhang, Chen, Lin, Sun, Song, Zhao, Xu, Dou, Wang, and Wen}]{UserBehaviorSim}
Wang, L.; Zhang, J.; Yang, H.; Chen, Z.-Y.; Tang, J.; Zhang, Z.; Chen, X.; Lin, Y.; Sun, H.; Song, R.; Zhao, X.; Xu, J.; Dou, Z.; Wang, J.; and Wen, J.-R. 2025.
\newblock User Behavior Simulation with Large Language Model-based Agents.
\newblock \emph{ACM Trans. Inf. Syst.}, 43(2).

\bibitem[{Wang et~al.(2017)Wang, Fu, Fu, and Wang}]{DCN}
Wang, R.; Fu, B.; Fu, G.; and Wang, M. 2017.
\newblock Deep \& Cross Network for Ad Click Predictions.
\newblock In \emph{Proceedings of the ADKDD'17}, ADKDD'17. New York, NY, USA: Association for Computing Machinery.
\newblock ISBN 9781450351942.

\bibitem[{Wang et~al.(2021)Wang, Shivanna, Cheng, Jain, Lin, Hong, and Chi}]{DCNv2}
Wang, R.; Shivanna, R.; Cheng, D.; Jain, S.; Lin, D.; Hong, L.; and Chi, E. 2021.
\newblock DCN V2: Improved Deep \& Cross Network and Practical Lessons for Web-scale Learning to Rank Systems.
\newblock In \emph{Proceedings of the Web Conference 2021}, WWW '21, 1785–1797. New York, NY, USA: Association for Computing Machinery.
\newblock ISBN 9781450383127.

\bibitem[{Xi et~al.(2024)Xi, Liu, Lin, Cai, Zhu, Zhu, Chen, Tang, Zhang, and Yu}]{TowardsOpen-WorldRec}
Xi, Y.; Liu, W.; Lin, J.; Cai, X.; Zhu, H.; Zhu, J.; Chen, B.; Tang, R.; Zhang, W.; and Yu, Y. 2024.
\newblock Towards Open-World Recommendation with Knowledge Augmentation from Large Language Models.
\newblock In \emph{Proceedings of the 18th ACM Conference on Recommender Systems}, RecSys '24, 12–22. New York, NY, USA: Association for Computing Machinery.
\newblock ISBN 9798400705052.

\bibitem[{Xu et~al.(2023)Xu, Li, Ha, Guo, Ma, Liu, Chen, and Zhu}]{xu2023neural}
Xu, W.; Li, S.; Ha, M.; Guo, X.; Ma, Q.; Liu, X.; Chen, L.; and Zhu, Z. 2023.
\newblock Neural node matching for multi-target cross domain recommendation.
\newblock In \emph{2023 IEEE 39th International Conference on Data Engineering (ICDE)}, 2154--2166. IEEE.

\bibitem[{Xu et~al.(2024{\natexlab{a}})Xu, Wu, Liang, Han, Ning, Shi, Lin, and Zhang}]{xu2024slmrec}
Xu, W.; Wu, Q.; Liang, Z.; Han, J.; Ning, X.; Shi, Y.; Lin, W.; and Zhang, Y. 2024{\natexlab{a}}.
\newblock SLMRec: Distilling large language models into small for sequential recommendation.
\newblock \emph{arXiv preprint arXiv:2405.17890}.

\bibitem[{Xu et~al.(2024{\natexlab{b}})Xu, Wu, Wang, Ha, Ma, Chen, Han, and Yan}]{RethinkingCDS}
Xu, W.; Wu, Q.; Wang, R.; Ha, M.; Ma, Q.; Chen, L.; Han, B.; and Yan, J. 2024{\natexlab{b}}.
\newblock Rethinking Cross-Domain Sequential Recommendation under Open-World Assumptions.
\newblock In \emph{Proceedings of the ACM Web Conference 2024}, WWW '24, 3173–3184. New York, NY, USA: Association for Computing Machinery.
\newblock ISBN 9798400701719.

\bibitem[{Zhai et~al.(2024)Zhai, Liao, Liu, Wang, Li, Cao, Gao, Gong, Gu, He, Lu, and Shi}]{ActionsSpeakLouder}
Zhai, J.; Liao, L.; Liu, X.; Wang, Y.; Li, R.; Cao, X.; Gao, L.; Gong, Z.; Gu, F.; He, J.; Lu, Y.; and Shi, Y. 2024.
\newblock Actions speak louder than words: trillion-parameter sequential transducers for generative recommendations.
\newblock In \emph{Proceedings of the 41st International Conference on Machine Learning}, ICML'24. JMLR.org.

\bibitem[{Zhang et~al.(2025)Zhang, Li, Long, Zhang, Lin, Yang, Xie, Yang, Liu, Lin, Huang, and Zhou}]{Qwen3Embed}
Zhang, Y.; Li, M.; Long, D.; Zhang, X.; Lin, H.; Yang, B.; Xie, P.; Yang, A.; Liu, D.; Lin, J.; Huang, F.; and Zhou, J. 2025.
\newblock Qwen3 Embedding: Advancing Text Embedding and Reranking Through Foundation Models.
\newblock arXiv:2506.05176.

\bibitem[{Zhao et~al.(2024)Zhao, Fan, Li, Liu, Mei, Wang, Wen, Wang, Zhao, Tang, and Li}]{RecInLLMEra}
Zhao, Z.; Fan, W.; Li, J.; Liu, Y.; Mei, X.; Wang, Y.; Wen, Z.; Wang, F.; Zhao, X.; Tang, J.; and Li, Q. 2024.
\newblock Recommender Systems in the Era of Large Language Models (LLMs).
\newblock \emph{IEEE Transactions on Knowledge and Data Engineering}, 36(11): 6889--6907.

\bibitem[{Zhou et~al.(2019)Zhou, Mou, Fan, Pi, Bian, Zhou, Zhu, and Gai}]{DIEN}
Zhou, G.; Mou, N.; Fan, Y.; Pi, Q.; Bian, W.; Zhou, C.; Zhu, X.; and Gai, K. 2019.
\newblock Deep interest evolution network for click-through rate prediction.
\newblock In \emph{Proceedings of the Thirty-Third AAAI Conference on Artificial Intelligence and Thirty-First Innovative Applications of Artificial Intelligence Conference and Ninth AAAI Symposium on Educational Advances in Artificial Intelligence}, AAAI'19/IAAI'19/EAAI'19. AAAI Press.
\newblock ISBN 978-1-57735-809-1.

\bibitem[{Zhou et~al.(2018)Zhou, Zhu, Song, Fan, Zhu, Ma, Yan, Jin, Li, and Gai}]{DIN}
Zhou, G.; Zhu, X.; Song, C.; Fan, Y.; Zhu, H.; Ma, X.; Yan, Y.; Jin, J.; Li, H.; and Gai, K. 2018.
\newblock Deep Interest Network for Click-Through Rate Prediction.
\newblock In \emph{Proceedings of the 24th ACM SIGKDD International Conference on Knowledge Discovery \& Data Mining}, KDD '18, 1059–1068. New York, NY, USA: Association for Computing Machinery.
\newblock ISBN 9781450355520.

\end{thebibliography}

\appendix
\section{Appendix}
\subsection{Details of KSF and ESA}
The core pseudocode of ESFNet serving as KSF and attention-based architecture serving as ESA are given in Listings \ref{ESFNetPseudocode} and \ref{ESAPseudocode}, where the scaling factor $\kappa$ is set as 2.
\begin{listing}[ht]
\caption{The ESFNet Pseudocode}%
\label{ESFNetPseudocode}%
\begin{lstlisting}[language=python]
def ESFNet(features_input, gate_input, block_dims):

 out_unit = len(block_dims)
 input_dim = gate_input.size(1)

 block_gate_input = gate_input.detach()
 gate_layer = ReLU(
    matmul(block_gate_input, W1) + b1
    )
 block_gate = 2.0 * Sigmoid(
    matmul(gate_layer, W2) + b2
    )

 block_gate_splits = split(
    block_gate, 
    dim=1, 
    size=1
    ) 
 feature_splits = split(
    features_input, 
    dim=1, 
    sizes=block_dims
    ) 

 gated_features = []
 for f, g in zip(feature_splits, block_gate_splits):
    gated_features.append(f * g)
 return concat(gated_features, dim=1)
\end{lstlisting}
\end{listing}
% \section{Parameter Sensitivity Analysis}
\begin{listing*}[ht]%
\caption{The Attention-Base Architecture Pseudocode}%
\label{ESAPseudocode}%
\begin{lstlisting}[language=python]
def compress_and_align_features(u_vec, i_vec, hist_emb, item_emb):
    compressed_u = mlp_u(u_vec).view()
    compressed_i = mlp_i(i_vec).view()

    user_behavior = mean(hist_emb, dim=1).view(bs, -1)
    query = concat([user_behavior, item_emb], dim=-1)

    att_u = cross_attention(compressed_u, query, W_q_u, W_k_u, W_v_u)  
    att_i = cross_attention(compressed_i, query, W_q_i, W_k_i, W_v_i)

    combined = concat([att_u, att_i], dim=1)
    attn_output, _ = multihead_attention(combined, combined, combined)

    flat = reshape(attn_output, [bs, -1])
    dens_vec = output_proj(flat)

    return dens_vec
\end{lstlisting}
\end{listing*}

\begin{table*}[t]
\centering
\setlength{\tabcolsep}{1mm}
\fontsize{9}{\baselineskip}\selectfont
\begin{tabular}{l *{12}{c}}
\toprule
 & \multicolumn{6}{c}{MovieLens-1M} & \multicolumn{6}{c}{Amazon-Book} \\
\cmidrule(lr){2-7} \cmidrule(lr){8-13}
Backbone
& \multicolumn{3}{c}{AUC $\uparrow$} & \multicolumn{3}{c}{LogLoss $\downarrow$} 
& \multicolumn{3}{c}{AUC $\uparrow$} & \multicolumn{3}{c}{LogLoss $\downarrow$} \\
\cmidrule(lr){2-4} \cmidrule(lr){5-7} \cmidrule(lr){8-10} \cmidrule(lr){11-13}
&{\makecell[c]{KSER \\ w/o \\ ESFNet}} & {\makecell[c]{KSER \\ w/o \\ ESA}}  &KSER 
&{\makecell[c]{KSER \\ w/o \\ ESFNet}} & {\makecell[c]{KSER \\ w/o \\ ESA}}  &KSER 
&{\makecell[c]{KSER \\ w/o \\ ESFNet}} & {\makecell[c]{KSER \\ w/o \\ ESA}} &KSER  
&{\makecell[c]{KSER \\ w/o \\ ESFNet}} & {\makecell[c]{KSER \\ w/o \\ ESA}}  &KSER \\
\midrule
DIN     &0.78268  &0.78343  &\textbf{0.78449}  
        &0.56020  &0.56473  &\textbf{0.55554}  
        &0.82928  &0.82889  &\textbf{0.83031}  
        &0.49535  &0.49431  &\textbf{0.49526}   \\
DCNv1   &0.78200  &\textbf{0.78302}  &0.78257   
        &0.55180  &0.55475  &\textbf{0.55161} 
        &0.83072  &0.83256  &\textbf{0.83282} 
        &0.49371  &0.49052  &\textbf{0.48965}  \\
DCNv2   &0.78134  &0.78291  &\textbf{0.78428}  
        &0.55553  &0.55343  &\textbf{0.55074}  
        &0.83033  &0.83220  &\textbf{0.83270}  
        &0.49695  &0.49011  &\textbf{0.48934}   \\
DIEN    &0.78266  &0.78069  &\textbf{0.78387}  
        &0.56020  &0.56075  &\textbf{0.55567}  
        &0.82865  &0.81097  &\textbf{0.82942}  
        &0.49585  &0.53863  &\textbf{0.49403}   \\
DeepFM  &0.77730  &0.77330  &\textbf{0.77901} 
        &0.56098  &0.56556  &\textbf{0.55517}  
        &0.83177  &0.83022  &\textbf{0.83190}  
        &0.49161  &0.49395  &\textbf{0.49150}   \\
FiBiNet &0.78213  &0.78303  &\textbf{0.78368}  
        &0.55214  &0.55210  &\textbf{0.55088}   
        &0.83166  &0.83205  &\textbf{0.83235}   
        &0.49219  &0.49075  &\textbf{0.49093}   \\
AutoInt &0.78024  &0.77751  &\textbf{0.78321}   
        &0.55477  &0.55876  &\textbf{0.55213}   
        &0.83090  &0.82932  &\textbf{0.83113}   
        &0.49328  &0.49484  &\textbf{0.49297}   \\
FiGNN   &0.78132  &0.77946  &\textbf{0.78295}   
        &0.55485  &0.55953  &\textbf{0.55305}   
        &0.83087  &0.83100  &\textbf{0.83153}   
        &\textbf{0.49350}  &0.49265  &0.49384   \\
xDeepFM &0.77961  &0.77667  &\textbf{0.78345}   
        &0.55657  &0.55789  &\textbf{0.55215}   
        &0.83177  &0.83187  &\textbf{0.83193}   
        &0.49798  &0.49144  &\textbf{0.49057}   \\
\bottomrule
\end{tabular}
\caption{Results of the ablation experiments for KSER with the extractor-only training strategy.}
\label{appendices_Tab_01}
\end{table*}
\subsection{Extractor-Only Training Ablation Studies}
\begin{figure}[ht]
\centering
\includegraphics[width=0.8\columnwidth]{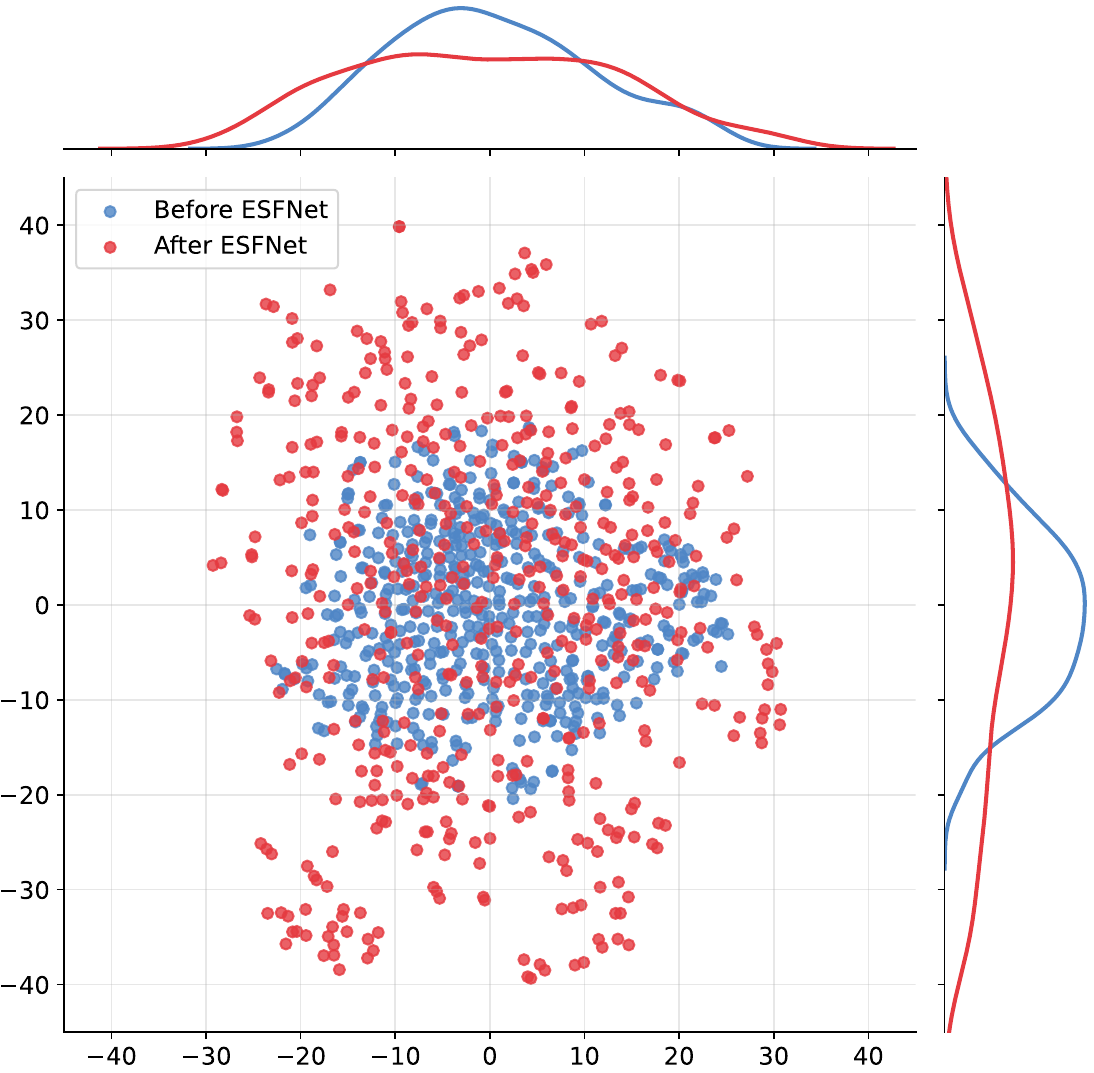} 
\caption{In extractor-only training strategy, embedding vectors before and after passing through ESFNet using T-SNE.}
\label{esfnet-extractor}
\end{figure}

\begin{figure}[ht]
\centering
\includegraphics[width=\columnwidth]{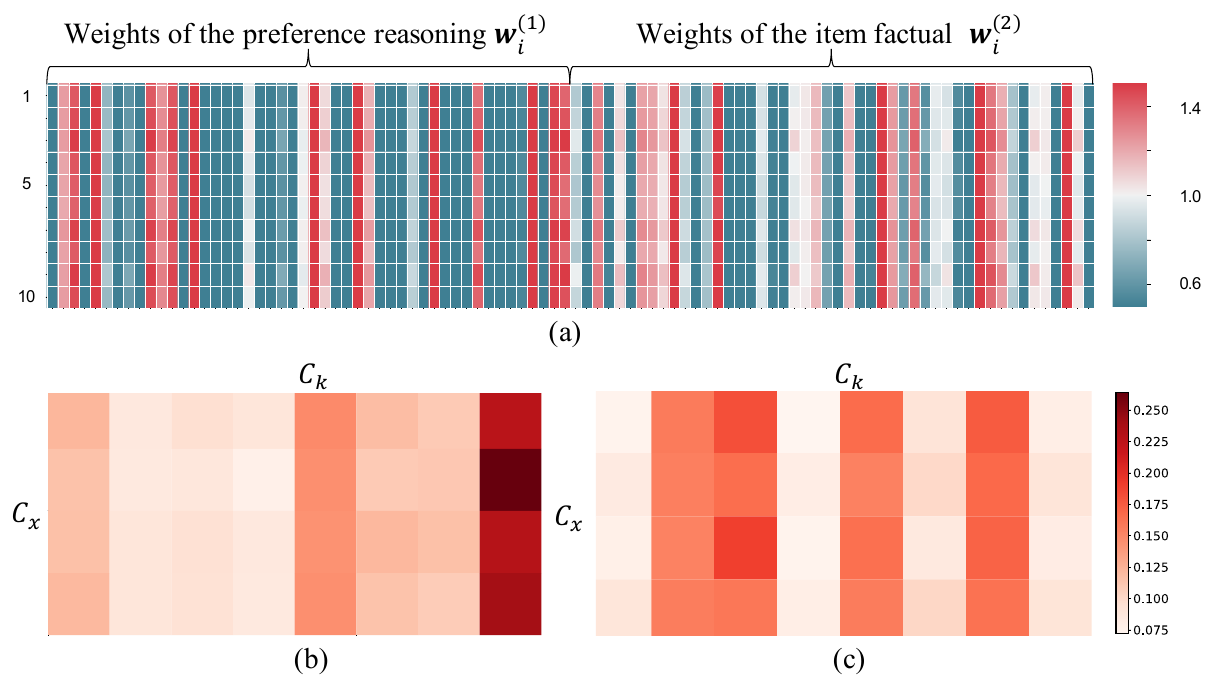} 
\caption{(a) In extractor-only training strategy, the weights from the gate neural unit for preference reasoning and item factual; (b) Attention score of the user preference filed in cross attention; (c) Attention score of the item factual filed in cross attention.}
\label{attscoreesfnetweight-extractor}
\end{figure}

In this section, the ablation experiments in extractor-only training strategy are conducted. The performance results are given in Table \ref{appendices_Tab_01}. Compared with the complete KSER framework with the extractor-only training strategy, KSER (w/o ESFNet) and KSER (w/o ESA) exhibit performance degradation, which demonstrates the necessity of ESFNet and ESA. Similar to the all-parameters training strategy, 512 samples are randomly chosen to visualize their embedding vectors (before and after pass- ing through ESFNet) in Fig. \ref{esfnet-extractor} using T-SNE for dimensionality reduction, where the knowledge embeddings after passing through ESFNet become more discriminative and the distributions become smoother. Compared with the all-parameters training strategy, the knowledge embeddings seem to be more discriminative, probably because the backbone model is frozen and then cannot compensate for the homogeneous and redundant knowledge embeddings. This training strategy forces the ESFNet to select more beneficial knowledge and filter out the homogeneous and redundant information, which can be verified by the the weights from the gate neural unit shown in Figs. \ref{attscoreesfnetweight-extractor}(a). The variance of weights for each sample increases significantly compared with the all-parameters training strategy. It can be observed the polarization of weights. Besides, the attention scores between $E(\textbf{x}_i)$ chunks and $\breve{\boldsymbol{k}}_{i}^{(j)}$ chunks of the user preference field and the  item factual field in ESA module are shown in Figs. \ref{attscoreesfnetweight-extractor}(b) and (c). In the extractor-only training strategy, the relevant information from $\bar{\boldsymbol{k}}_{i}^{(j)}$ with respect to $E(\textbf{x}_i)$ can be effectively extracted.
\end{document}